%% file: main.tex
\documentclass[a4paper,11pt]{article}
\usepackage{jcappub}
\usepackage{lineno}

\input{macros.tex}

\usepackage{lettrine}
\usepackage{graphicx}
\usepackage{booktabs}

\title{\begin{center} Large scale structure prior knowledge in the dark siren method \end{center}}

\author{\begin{center} Charles Dalang, Bartolomeo Fiorini and Tessa Baker \end{center}}
\affiliation{\begin{center}\textit{School of Mathematical Sciences, Queen Mary University of London, \\ Mile End Road, E1 4NS, London, United Kingdom}\end{center}}
\affiliation{\begin{center} \textit{Institute of Cosmology and Gravitation, University of Portsmouth, \\
Burnaby Road, Portsmouth PO1 3FX, United Kingdom}\end{center}}

\emailAdd{c.dalang@qmul.ac.uk}
\emailAdd{tessa.baker@port.ac.uk}

\abstract{Gravitational wave dark sirens are a powerful tool for cosmology and inference of compact object population hyperparameters. They allow for a measurement of the luminosity distance to the source, but not their redshift. Galaxy catalogues in the source localization volume can be used to infer the redshift of the source in a statistical manner. Catalogues are, however, limited by their incompleteness, which can be significant at redshifts corresponding to current GW events. In this work, we detail how to implement in practice variance completion, a novel galaxy completion method which uses knowledge of the large scale structure to optimize the potential of dark sirens analyses. We compress the prediction for the missing number of galaxies into a ratio between the predictions of variance completion and the standard homogeneous completion method. This ratio format can be easily incorporated into existing line of sight computations used in dark sirens software; we demonstrate this procedure using the GLADE+ galaxy catalogue and the {\tt gwcosmo} software package. We discuss the robustness of the method, and apply it to well-localized event GW190814 as a proof of concept. Finally, we apply the method to data from the third observing run of LIGO-Virgo-KAGRA, finding that it yields results that are consistent with homogeneous completion. We also discuss the prospects for an improvement if the GW localization volume shrinks. }

\begin{document}
\maketitle
\flushbottom

\section{Introduction}

After centuries of astronomy based on the electromagnetic spectrum, it has now been almost a decade since the beginning of the era of gravitational wave astronomy. As we will see in this work, the topic of gravitational wave (GW) cosmology sits well at the interplay of both fields. One of the major objectives of GW cosmology is to obtain independent measurements of cosmological parameters, and to infer hyperparameters of the compact object population. 

In a similar time frame, the so-called \textit{Hubble tension} has emerged as a topic of interest. The lower value of $H_0 = (67.4 \pm 0.5)$km\,s$^{-1}$Mpc$^{-1}$ obtained from the cosmic microwave background (CMB) by assuming a cosmological model \cite{Planck:2018vyg} is in tension at about $(4-6)\sigma$ with local measurements, which cluster around $H_0 = 74$ km\,s$^{-1}$Mpc$^{-1}$ \cite{Riess:2019cxk,Riess:2021jrx,Pesce:2020xfe,H0LiCOW:2019pvv}. It has been shown to be incredibly difficult to solve the Hubble tension by invoking some form of new physics, without also disrupting constraints from the CMB, big bang nucleosynthesis and large scale structure \cite{Schoneberg:2021qvd,DiValentino:2021izs}. At the same time, the tension seems to be equally robust to all known sources of systematics \cite{Brout:2021mpj}. GWs offer a promising avenue to add a completely independent measurement of the Hubble constant \cite{Schutz:1986gp}, which is however not without challenges \cite{Pierra:2023deu}. 

GW methods are attractive because the luminosity distance is directly encoded in the waveform of compact binary coalesences. The tricky step in inferring a Hubble constant in this scenario is to extract the redshift of the source. In rare cases, the GW source can also emit some electromagnetic radiation, which can be used to pinpoint the host galaxy and extract the redshift spectroscopically. These systems are called bright sirens and one event has been detected so far, leading to a Hubble constant measurement of $H_0=70.0^{+12.0}_{-8.0}$km\,s$^{-1}$Mpc$^{-1}$ sitting nicely in the middle of the two $H_0$ schools \cite{LIGOScientific:2017adf}. In the future, we hope to get more of these bright sirens \cite{Sathyaprakash:2009xt}, which also requires to model electromagnetic selection effects \cite{Mancarella:2024qle}. Meanwhile, there are about $100$ publicly available dark sirens, i.e.\,those without electromagnetic counterparts, which can be used to study cosmology using different statistical techniques. On one hand, observations of the redshifted masses, and knowledge of features of the source frame distribution of compact object masses can be used to extract the redshift of the source in a statistical way \cite{Ezquiaga:2022zkx,Mastrogiovanni:2021wsd,Finke:2021eio,Mancarella:2021ecn, Mancarella:2022cnu, Iacovelli:2022tlw,Karathanasis:2022rtr, Fung:2023yyq}. On the other hand, a galaxy catalogue overlapping the GW localization volume of the source can be used to reconstruct the Hubble constant statistically \cite{Gray:1,Gray:2,Gray:2023wgj}. Both of these methods can now be used simultaneously to infer the Hubble constant and compact object population hyperparameters jointly, using the publicly available codes \texttt{gwcosmo}, \texttt{icarogw} and \texttt{CHIMERA} \cite{Gray:2023wgj,Mastrogiovanni:2023emh,Borghi:2023opd}.  

A major difficulty in using galaxy catalogues to infer the redshift of dark sirens is related to the completeness of galaxy surveys. As surveys are often magnitude-limited,   the completeness drops with redshift; for many currently available surveys, a low level of completeness is reached whilst still at redshifts of interest for GW science. Moreover, completeness is highly anisotropic due to the different footprints of surveys, taken with telescopes with various different properties. 

Several methods have been introduced in the literature to mitigate incompleteness, including homogeneous, multiplicative and variance completion \cite{LIGOScientific:2018gmd, Finke:2021aom, Dalang:2023ehp}. Homogeneous completion is the simplest technique, and the method currently implemented in \texttt{gwcosmo} and \texttt{icarogw}. It assumes that the out-of-catalogue galaxies are distributed homogeneously in comoving volumes. While simple, this method ignores the tendency of gravity to cluster galaxies in a way which is well understood. To compensate for this, multiplicative completion was introduced in \cite{Finke:2021aom}, assuming a proportionality factor between the galaxies that are in and those that are out of the catalogue. This has been shown, however, to overestimate the amplitude of structure at low completeness. Additionally, it lacks a robust physical foundation. Conversely, variance completion has been introduced to make use of the fact that galaxies are clustered in a way which can be estimated, so as to fill in missing galaxies, where we expect them to be, from large scale structure knowledge. This method was tested on simulated data in two dimensions and shown to outperform the other methods, so long as the structure of the catalogued galaxies represents sufficiently well the true distribution \cite{Dalang:2023ehp}. Note that a sophisticated completion method taking into account large scale structure and the magnitude distribution of sources is introduced in \cite{Leyde:2024tov}. Alternatively, one can focus on events which have very good catalogue support by selecting GW events which overlap well with a volume limited sample of galaxies from DES \cite{DES:2019ccw} or DESI \cite{DESI:2023fij}.

In this work we build upon the theoretical foundations of \cite{Dalang:2023ehp}, detailing how to implement variance completion on real data. We implement it on the GLADE+ galaxy survey \cite{Dalya:2021ewn}, as an example. We discuss a number of practical adaptations to be made, for example, to incorporate redshift uncertainties in the method. We introduce a ratio function, which can easily be incorporated into the formalism of \texttt{gwcosmo}. We discuss and marginalize over the cosmology dependence where relevant. We analyze GW190814, a well-localized GW event with homogeneous and variance completion and discuss the differences on the Hubble constant posterior. Finally, we apply our variance-completed line of sight priors to dark siren events from the third observing run of LIGO-Virgo-KAGRA (LVK) with sufficient signal to noise ratio (SNR), and show how the method affects the posterior distribution of the Hubble constant. The method works particularly well for well-localized events, of which we expect to see more, as GW detectors improve.

\section{Methodology} \label{sec:Methodology}
In this section, we describe in detail how we implement in practice the formalism of variance completion on real data. We start by describing the line of sight redshift prior in Sec.\,\ref{subsec:LOS_prior} to introduce how the incompleteness of galaxy catalogues affects this. We then describe how to discretize the galaxy field in voxels and how we use a magnitude threshold map to identify voxels of similar completeness in Sec.\,\ref{subsec:Voxalization} and \ref{subsec:Completeness}. We then briefly review homogeneous and variance completion in Sec.\,\ref{subsec:Homogeneous_Completion} and \ref{subsec:Variance_Completion} and detail how to compute large scale structure ingredients required for these methods in Sec.\,\ref{subsec:average_variance}. We explain how to implement the method in a code such as \texttt{gwcosmo} in Sec.\,\ref{subsec:gwcosmo} and discuss the cosmology dependence of the method in Sec.\,\ref{subsec:Cosmology_Dependence} 

\subsection{Line of sight redshift prior}\label{subsec:LOS_prior}
One required ingredient for the dark siren method, which is the essence of this work, is the line of sight redshift prior. It encodes the probability density that the GW was emitted from a particular redshift $z$ and direction $\bs{\hat{n}}$ on the sky. GWs come from binary compact objects, which live inside galaxies. It is therefore clear that this probability density should depend on the galaxy field. In principle, we would like to weight each galaxy according to their probability to host a GW event. Unfortunately, this quantity is difficult (if not impossible) to model. Alternatively,  we can expect this quantity to be correlated with some galaxy indicators, such as luminosity, mass or star formation rate. For the completion method that we implement in this work, we work with luminosity weighting for all galaxies. 
We will then show how to combine this completion method with catalogues which use some form of galaxy weighting for the galaxies in the catalogue in Sec.\,\ref{subsec:gwcosmo}. 

One can write that the probability density that a GW was emitted from redshift $z$ and direction $\bs{\hat{n}}$ is proportional to the weighted galaxy number density
\begin{align}
p(z,\bs{\hat{n}}) = 
\frac{\frac{\dd n\e{g}}{\dd z \dd \Omega} (z,\bs{\hat{n}})}{\int \dd z \dd \Omega \,  \frac{\dd n\e{g}}{\dd z \dd \Omega} (z,\bs{\hat{n}})}\,,
\end{align}
where $\dd n\e{g}/(\dd z \dd \Omega)(z,\bs{\hat{n}})$ denotes the galaxy number density per redshift bin $\dd z$ and solid angle $\dd \Omega$. The total number of galaxies in the volume under consideration reads
\begin{align}
N\e{g} = \int \dd z \dd \Omega \,  \frac{\dd n\e{g}}{\dd z \dd \Omega} (z,\bs{\hat{n}})
\end{align}
One should distinguish between the galaxies which are observed in the catalogue $n\e{c}$ from those which are missed $n\e{m}$. Those should sum up to the total number of galaxies $n\e{g} = n\e{c}+n\e{m}$ on any volume $\dd \Omega \dd z $. Therefore,
\begin{align}
p(z,\bs{\hat{n}}) = p\e{c}(z,\bs{\hat{n}}) + p\e{m}(z,\bs{\hat{n}})\,, \label{eq:LOS}
\end{align}
with 
\begin{align}
p\e{c}(z,\bs{\hat{n}})&  = \frac{1}{N\e{g}}  \frac{\dd n\e{c}}{\dd z \dd \Omega}(z,\bs{\hat{n}})\,, \\
p\e{m}(z,\bs{\hat{n}}) & =  \frac{1}{N\e{g}} \frac{\dd n\e{m}}{\dd z \dd \Omega} (z,\bs{\hat{n}})\,,\label{eq:dn_m/dzdOmega}
\end{align}
The galaxies in the catalogue are well localized in direction but less so in redshift space, especially for photometric galaxies. Hence, we include the uncertainty on redshift but neglect the uncertainty on direction. The probability density can be expressed as a sum over the $N\e{c}$ catalogued galaxies
\begin{align}
p\e{c}(z,\bs{\hat{n}}) = \frac{1}{N\e{g}} \sum_{i=1}^{N\e{c}} \, p(z;\bar{z}^i,\sigma^i) \delta^2(\bs{\hat{n}}- \bs{\hat{n}}^i)\,.
\end{align}
For the out-of-catalogue galaxies in Eq.\,\eqref{eq:dn_m/dzdOmega}, we will estimate $\frac{\dd n\e{m}}{\dd z \dd \Omega}$ using homogeneous or variance completion, that we describe in Sec.\,\ref{subsec:Homogeneous_Completion} and \ref{subsec:Variance_Completion}. 

\subsection{Discretization} \label{subsec:Voxalization}

\begin{figure}[ht]
\centering
\includegraphics[width=0.7\textwidth]{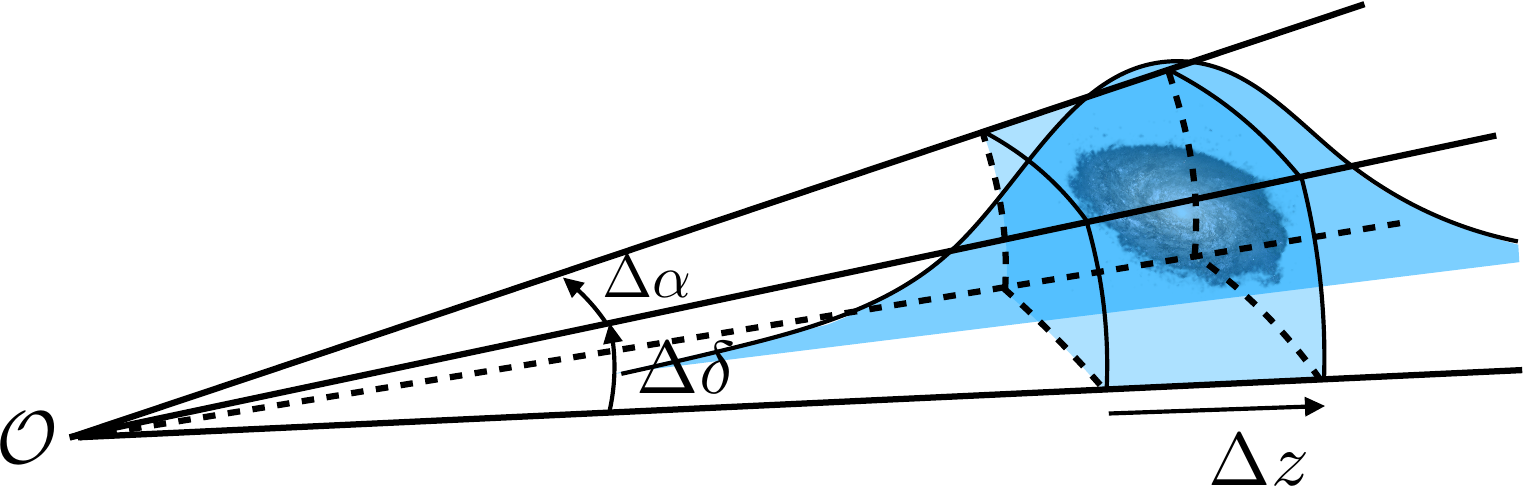}
\caption{We sketch the contribution of 1 galaxy to a voxel. A galaxy can contribute a fraction of itself to several voxels. This is typically the case for photometric redshift bins, while spectroscopic redshift contribute in general to only 1 voxel.} 
\label{fig:voxel_galaxy}
\end{figure}

For the purpose of implementing variance completion, it will prove useful to split the galaxies into voxels, i.e.\,the 3 dimensional counterpart of pixels. The number of galaxies in a volume $\mathcal{V}\e{vox}^i$, defined by its redshift and solid angle support, relate to the galaxy densities via an integral over the voxel volume $\mathcal{V}\e{vox}^i$ 
\begin{align}
n^i\e{g} = \int_{\mathcal{V}\e{vox}^i} \dd \Omega \dd z \frac{\dd n\e{g}}{\dd \Omega \dd z} (z,\bs{\hat{n}})\,,
\end{align}
where the index $i$ indicates the voxel number. A similar relation applies to $n\e{c}$ and $n\e{m}$. We sketch the contribution of a galaxy to a voxel in Fig.\,\ref{fig:voxel_galaxy}.
One galaxy can spread over multiple voxels depending on its redshift uncertainty. This allows for $n^i\e{c}$ to capture these redshift uncertainties. When combining information from different surveys, some galaxy redshifts have widely different error bars. Some are photometric, others are spectroscopic. We take the redshift error bars $\sigma_z^i$ into account by assuming that the redshift posteriors are well modeled by Gaussian PDFs, which is not necessarily a good approximation. Hence, the number of galaxies in each voxel $n^i\e{c}$ reads
\begin{align}
n^i\e{c}(z,\bs{\hat{n}}) = \sum_j \int_{z_i-\delta z/2}^{z_i+\delta z/2} \dd z \, p(z;\bar{z}^j,\sigma_z^j)\,,
\end{align}
where the index $j$ runs over all catalogued galaxies in the pixel corresponding to the direction $\bs{\hat{n}}$ and $p(z;\bar{z}^j,\sigma_z^j)$ are normal distributions centered on $\bar{z}^j$ with standard deviation $\sigma_z^j$. Therefore, the number of galaxies in each voxel does not need to be an integer. 

We shall fix the angular size of voxels throughout each analysis. We shall also fix redshift bins $\delta z$ implying that voxels change volume with redshift. The comoving volume of voxels at redshift $z$ reads 
\begin{align}
\mathcal{V}\e{vox}(z) = \int_{z-\delta z/2}^{z+\delta z/2} \dd z \, \frac{\dd V\e{c}}{\dd z}(z)\,, \label{eq:Volume}
\end{align}
where $\dd V\e{c}/(\dd z\dd\Omega) = c\,r^2(z)/H(z)$, such that $r(z)$ is the comoving distance and $H(z)$ the Hubble rate at redshift $z$. Those are given, for a flat $\Lambda$CDM cosmology by
\begin{align}
r(z) = \int_0^z \frac{\dd z'\,c}{H(z')}\,, \label{eq:comoving_distance}
\end{align}
and
\begin{align}
H(z) = H_0 \sqrt{\Omega\e{m0}(1+z)^3 + (1-\Omega\e{m0})}\,,
\end{align} 
where $H_0$ indicates the present day Hubble constant and $\Omega_{m0}$ denotes the present day energy fraction of matter in the Universe. This is the first instance in the method, which requires a cosmological model. We discuss the cosmology dependence of the entire procedure in Sec.\,\ref{subsec:Cosmology_Dependence}. For a fixed pixel size, two effects affect the volume of a voxel as redshift grows. The depth of the voxel decreases slightly, while the width perpendicular to the line of sight grows. These two effects combine to a net increase in the voxel volume as redshift grows. 

We have identified a way to divide the galaxy field into voxels taking into account galaxy redshift uncertainties. In the following section, we discuss the completeness fraction as a way to identify voxels of similar completeness levels, which is a required ingredient of variance completion.

\subsection{Completeness fraction}\label{subsec:Completeness}

For homogeneous and variance completion, it is necessary to estimate the completeness fraction of galaxies as a function of direction and redshift. For homogeneous completion, it allows to estimate the amplitude of the out-of-catalogue contribution. For variance completion, it allows to determine voxels of similar completeness, on which the algorithm can apply.

\begin{figure}[ht]
\centering
\includegraphics[width=1.0\textwidth]{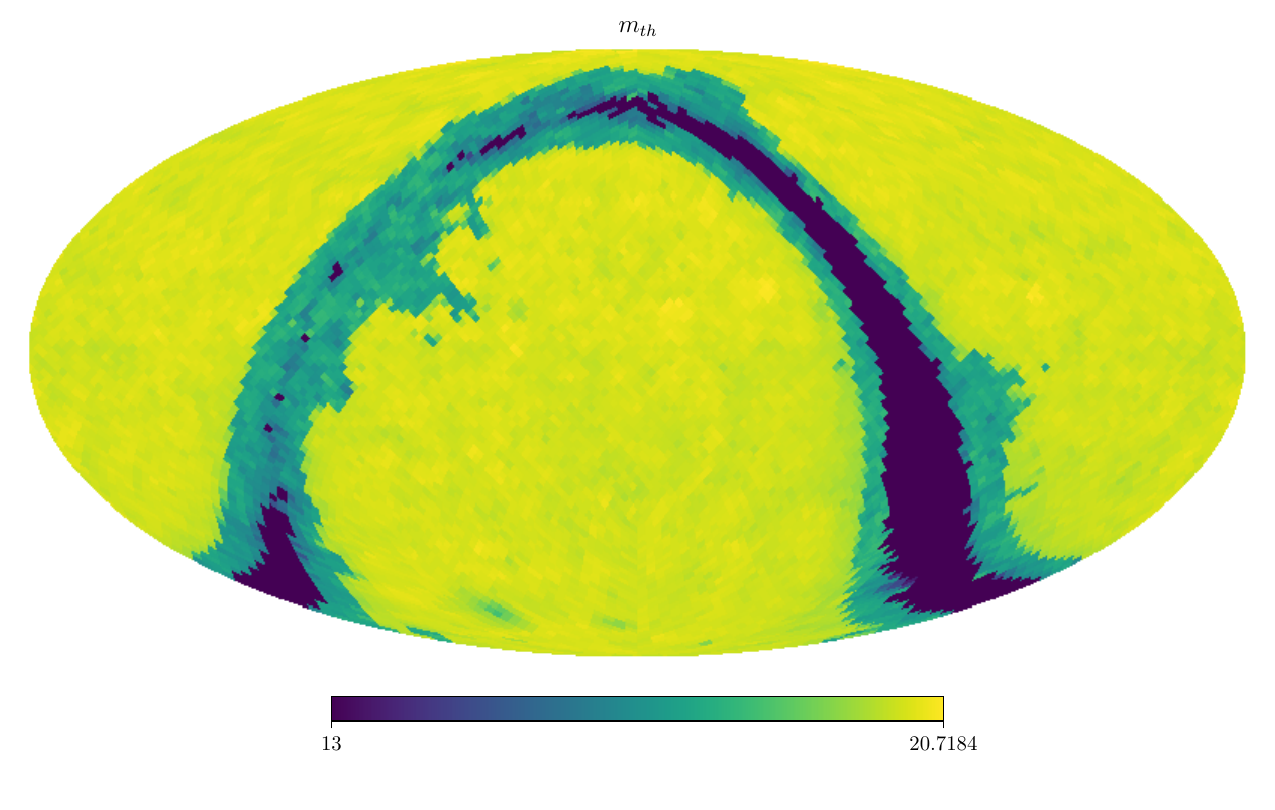}
\caption{Here, we show the magnitude threshold map $m\e{th}(\bs{\hat{n}})$ in the K-band for the GLADE+ galaxy catalogue for $n\e{side} = 32$, from which we can compute the completeness fraction in each voxel. Pixels for which $m\e{th}(\bs{\hat{n}})<13$ are plotted with the same color as $m\e{th}(\bs{\hat{n}}) = 13$. The minimum magnitude threshold determined is $ 10.378$, while pixels left blank have $m\e{th} = -\infty$. The higher the magnitude threshold, the better we estimate a pixel to have been observed.} 
\label{fig:mth_map}
\end{figure}

\begin{figure}[ht]
\centering
\includegraphics[width=1.0\textwidth]{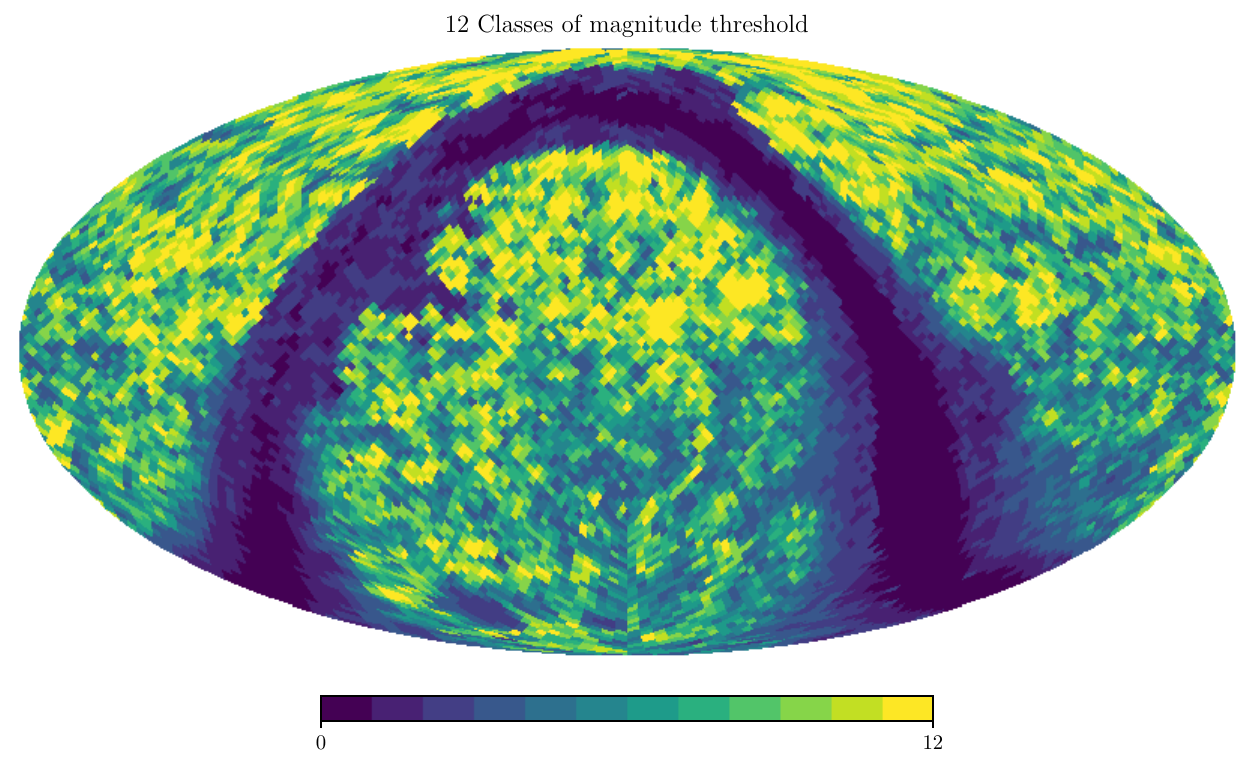}
\caption{Here, we show a map of the 12 classes of pixels with similar estimated completeness fractions $\hat{f}_\S$. This map is found by ranking pixels according to their magnitude threshold, which is presented in Fig.\,\ref{fig:mth_map}.}  
\label{fig:Ngroups}
\end{figure}

For a given galaxy catalogue, a magnitude threshold map $m\e{th}(\bs{\hat{n}})$ can be determined to establish groups of pixels with similar levels of completeness at each redshift \cite{Gray:2023wgj}. We call these groups, \textit{classes}. The estimate of the magnitude threshold is based on the median apparent magnitude of galaxies in the relevant (B- or K- band) band and returns $-\infty$ for empty pixels, i.e.\,those which have less galaxies than a certain threshold, fixed to 10 galaxies. This allows the magnitude threshold map to be relatively insensitive to outliers. The threshold of 10 galaxies per pixel prevents an estimate of an imprecise magnitude threshold from the median of the apparent magnitude from a too small sample of galaxies, which may propagate to the completeness. We order pixels as a function of $m\e{th}(\bs{\hat{n}})$ and split them in classes of 1024 voxels. For $n\e{side} = 32$, which divides the sphere in 12,288 equal area pixels of about $0.0010$ sr,  we count 12 classes of $1024$ \texttt{healpix} pixels \cite{Zonca2019, healpix:2005}. \footnote{\url{http://healpix.sourceforge.net}} We shall apply the variance completion algorithm on these different classes for the B-band. The first class ordered by completeness levels has a few pixels which are observed very deeply. This results in significant variations in the completeness levels among the pixels of this class and we cannot trust the results of variance completion. Likewise the last four classes ordered by completeness levels have significant variations in the magnitude threshold map. They span mostly the regions close to the Milky way and we cannot trust the results of variance completion if the completeness varies a lot from pixel to pixel in the same class. We discard these 5 classes to which we apply homogeneous completion. We note that for the 7 other classes, the difference between the maximum and minimum $m\e{th}(\bs{\hat{n}})$ is smaller than $0.25$ mag. In Fig.\,\ref{fig:mth_map}, we plot the magnitude threshold map. We also show in Fig.\,\ref{fig:Ngroups} a map of the 12 classes of pixels with similar completeness. The completeness is calculated as
\begin{align}
\hat{f}_\S = \frac{\int_{L(m\e{th}(\bs{\hat{n}}))}^{L\e{max}} \dd L \,L \phi(L) }{\int_{L\e{min}}^{L\e{max}}\dd L \,L \phi(L) }\,, \label{eq:Completeness_Fraction}
\end{align}
where the luminosity function is assumed to be a Schechter function \cite{Schechter:1976iz}
\begin{align}
\phi(L) = \frac{\phi_*}{L_*} \l(\frac{L}{L_*}\r)^{\alpha} \hbox{exp}\l( -L/L_*\r)\,,
\end{align}
with K-band values $L\e{*;K}$, $L\e{min;K}$ and $L\e{max;K}$ calculated from $M \e{*;K} = -23.39$, $M\e{min;K} = -27$ and $M\e{max;K}=-19$ and $\alpha_K=-1.09$ \cite{Kochanek:2000im}. For the B-band, we use $M\e{*;B} = -19.66$, $M\e{min;B} = -22$ and $M\e{max;B}=-16.5$ and $\alpha\e{B}=-1.21$ \cite{Norbert:2002}. The conversion from apparent magnitude $m$ to absolute magnitude $M$ is done via
\begin{align}
M(m) = m - 5 \log_{10}(d_L(z)) - 25 - k(z)\,, \label{eq:m_and_z_to_M}
\end{align}
with $k$-correction given by $k\e{K}(z) = -6 \log_{10}(1+z)$ for $z\leq 0.25 $ for the K-band\footnote{Note that we apply the variance completion algorithm to redshifts lower than this.} \cite{Kochanek:2000im} and $k\e{B}(z) = (z+6z^2)/(1+15 z^3)$ \cite{Norbert:2002} for the B-band. The luminosity distance as a function of redshift is obtained via
\begin{align}
d_L(z) = (1+z) r(z)\,,
\end{align}
where $r(z)$ is the comoving distance as a function of redshift, given in Eq.\,\eqref{eq:comoving_distance}.
The luminosities can be obtained from an absolute magnitude using
\begin{align}
L(M) = L_0 \cdot 10^{-2M/5}\,,\label{eq:M_to_L}
\end{align}
with $L_0 = 3.0128 \cdot 10^{28}$ W. One can use these relations to convert the apparent magnitude threshold to a luminosity threshold and compute the completeness fraction in each voxel from Eq.\,\eqref{eq:Completeness_Fraction}. This completeness fraction is required to apply homogeneous or variance completion, as detailed in the following two sections. 

\subsection{Homogeneous completion} \label{subsec:Homogeneous_Completion}
In this section, we review homogeneous completion, introduced in \cite{LIGOScientific:2018gmd}, on which variance completion builds up. In the previous section, we determined classes $\S$ of pixels which have similar completeness at fixed redshift. Given a region $\S$ of $N_k$ voxels with the same estimated completeness fraction $\hat{f}_\S\neq 0$, one can estimate the number of galaxies in each voxel as
\begin{align}
\hat{\bar{n}}\e{g} = \frac{\sum_{i=1}^{N_k} n\e{c}^i}{N_k \hat{f}_\S}\,.\label{eq:nhatg}
\end{align}
This comes from the expectation that over sufficiently many voxels, the completeness should satisfy $f_\S = \sum_{i=1}^{N_k} n^i\e{c}/\sum_{i=1}^{N_k} n^i\e{g}$. Homogeneous completion estimates the number of missing galaxies in each voxel as
\begin{align}
\hat{n}\e{m;hom}^i = \hat{\bar{n}}\e{g} (1-\hat{f}_\S) =\frac{1-\hat{f}_\S}{\hat{f}_\S}\cdot \frac{\sum_i^{N_k} n^i\e{c}}{N_k}\,, \label{eq:homogeneous_estimator}
\end{align}
where the completeness fraction can be estimated from the catalogue, as discussed in the previous section. The first expression is useful to estimate the number of missing galaxies even when $\hat{f}_\S=0$, in which case, the second expression is undetermined.

The advantage of this method is that it can be computed once for all the voxels in the relevant class $\S$ which have completeness $\hat{f}_\S$. In this sense, it is very simple. We approximate the number density by dividing the number of galaxies by the corresponding redshift bin $\Delta z^i$ and solid angle $\Delta \Omega^i$
\begin{align}
\frac{\dd n\e{m;hom}}{\dd z \dd \Omega} (z^i,\bs{\hat{n}}^i)\simeq \frac{\hat{n}^i\e{m;hom}}{\Delta z^i \Delta \Omega^i}\,,
\end{align}
The downside of this method is that it underestimates the amplitude of structure \cite{Dalang:2023ehp}. This does not make optimal use of the information brought into dark siren analyses by galaxy catalogues. In the next section, we review variance completion, which is designed to take into account large scale structure, to complete a galaxy catalogue.

\subsection{Variance Completion}\label{subsec:Variance_Completion}
In Ref.\,\cite{Dalang:2023ehp}, we presented \textit{variance completion}, a way to complete a galaxy catalogue which makes use of knowledge from large scale structure. This method is quite sophisticated on its own and the subject of an entire article \cite{Dalang:2023ehp}. Here, we give only a very crude reminder of the method and defer the details to the original paper \cite{Dalang:2023ehp}.

On top of the knowledge of the mean galaxy number density $\bar{n}\e{g}$, variance completion also requires knowledge of the standard deviation in the galaxy field $\sigma\e{g}$. Given these two ingredients, determining the number density of missing galaxies $n^i\e{m}$ boils down to minimizing a well-chosen function. It was shown that minimizing the following function with respect to all the $n^i\e{m}$ worked very well \cite{Dalang:2023ehp}:
\begin{align}
\mathcal{L}[n^1\e{m},n^2\e{m},\dots,n^{N_k}\e{m}] = A_0 \sum_{i=1}^{N_k} (n^i\e{m})^2 + B_k \Bigg[ \l( \frac{1}{N_k} \sum_{i=1}^{N_k} (n^i\e{m}+n^i\e{c})^2\r) - (\bar{n}\e{g}^2 + \sigma\e{g}^2)\Bigg]^2\,, \label{eq:Lagrangian}
\end{align}
where $A_0$, $B_k \in\mathbb{R}$ are real constants. All number densities $n^i\e{m}$, $n^i\e{c}$ are understood to belong to the same class $\S$ of $N_k$ pixels, at the same redshift $z$. Fixing the redshift ensures that the volume of each of the voxels involved in this sum is the same and that comparing number densities in each voxels of this class makes sense. Intuitively, the minimization of the first term ensures that only the minimum number of missing galaxies are introduced. The second term ensures that the standard deviation of the estimated galaxies $\hat{n}\e{g} = \hat{n}\e{m}+n\e{c}$ respects the expectation from the standard deviation in the galaxy field $\sigma\e{g}$.

The minimum can be found by setting the gradient of Eq.\,\eqref{eq:Lagrangian} to zero, i.e.\,$\bs{\nabla}\mathcal{L} = \bs{0}$. This results in a complicated set of nonlinear coupled equations, which can be linearized around an initial guess $n^i\e{m} =\bar{n}^i\e{m}+\delta n^i\e{m}$. This initial guess is provided by Eq.\,\eqref{eq:homogeneous_estimator} from homogeneous completion, i.e.\,$\bar{n}^i\e{m} \to n^i\e{m;hom} $. The linear system can be written in matrix form as $A \cdot \bs{\delta n}\e{m} = \bs{b}$, where the form of $A$ and $\bs{b}$ are given in \cite{Dalang:2023ehp} and depend on the initial guess $\bar{n}^i\e{m}$. The linear system can then be inverted to find $ \bs{\delta n}\e{m} = A^{-1}\bs{b}$. These solutions can then be fed into a better initial guess as $\bar{n}\e{m}^i \to\bar{n}\e{m}^i +\delta n\e{m}^i$. This iteration can be made until the solution converges to the desired accuracy. The values of $A_0$ and $B_k$ can be calibrated such that the total number of missing galaxies in $\S$ averages to the total expected number of galaxies from homogeneous completion in $\S$, i.e.\,$N_k \bar{n}\e{m;hom}$ within $1\%$. Finally, the variance completion estimators are given by $\hat{n}^i\e{m;var} = \bar{n}^i\e{m} +\delta n^i\e{m}$. We can approximate the number density by dividing by the redshift and solid angle volume of the voxels
\begin{align}
\frac{\dd n\e{m;var}}{\dd z \dd \Omega} (z^i,\bs{\hat{n}}^i)\simeq \frac{\hat{n}\e{m;var}^i}{\Delta z^i \Delta \Omega^i}
\end{align}
In the rest of this article, we aim to implement this method in practice to reconstruct the line of sight redshift prior (given in Eq.\,\eqref{eq:LOS}) in a way which is compatible with \texttt{gwcosmo}. Furthermore, we check whether homogeneous or variance completion gives a more informative posterior on the Hubble constant. In the following, we detail how the average and standard deviation in the galaxy field can be estimated.

\subsection{Average number density and standard deviation}\label{subsec:average_variance}
As hinted in the previous section, variance completion requires knowledge of the standard deviation in the number of galaxies per voxel $\sigma\e{g}$ on top of the average number of galaxies per voxel $\bar{n}\e{g}$, which is also an ingredient of homogeneous completion. In this subsection, we detail how they are determined. 

For the average number density of galaxies, one can use the results of \cite{Conselice:2016}, which provide a formula for the comoving number density as a function of cosmic time, which is however, cosmology dependent. Instead, we work with the average number density $\hat{\bar{n}}\e{g}$ calculated from Eq.\,\eqref{eq:nhatg}, determined from the catalogue itself. As we shall see, this quantity ends up canceling when implemented in \texttt{gwcosmo}. 

The variance in the galaxy number counts requires more work. It can be computed from the variance $\sigma^2(R)$ in the smoothed dark matter density contrast $\delta_s(z,\bs{x},R) $, where $R$ represents the smoothing scale.
\begin{align}
\sigma^2(R) & \equiv \langle \delta_s^2(z,\bs{x},R) \rangle - \underbrace{\langle \delta_s(z,\bs{x},R)\rangle^2 }_{=0}\,,
\end{align}
where the smoothed density contrast reads
\begin{align}
\delta_s(z,\bs{x},R) = \int_{\mathbb{R}^3} \dd^3 \bs{x}\, \delta\e{dm}(z,\bs{x'}) W_s(|\bs{x}'- \bs{x}|, R)\,.
\end{align}
There, the window function $W_s(|\bs{x'}-\bs{x}|,R)$ smooths the dark matter density contrast $\delta\e{dm}(z,\bs{x}) \equiv (\rho\e{dm}(z,\bs{x})-\bar{\rho}\e{dm}(z))/\bar{\rho}\e{dm}(z)$ on a scale $R$ around the position $\bs{x}$. The local dark matter energy density $\rho\e{dm}(z,\bs{x})$ averages on spacelike hypersurfaces of constant cosmic time defined by the cosmological redshift $z$ to $\bar{\rho}\e{dm}(z)$. In principle, the window function support domain should take the shape of the voxels. For simplicity, we assume that the voxels are equivalent in shape to spheres, 
and neglect any errors arising from not capturing their full shape. As we shall see, we only need an approximate estimate of the variance. For this purpose, we extract the redshift-dependent radius from the voxel volume $\mathcal{V}\e{vox}(z) = 4 \pi R^3(z)/3$, which can be calculated using Eq.\,\eqref{eq:Volume} and compute the corresponding window function as follows
\begin{align}
W_s(x, R(z)) = \begin{cases} \frac{3}{ 4 \pi R^3(z)} \,, & \hbox{if}\, x\leq R(z) \,,\\ 
0\,, & \hbox{otherwise.} \end{cases} 
\end{align}
Its Fourier transform reads
\begin{align}
\t{W}_s(k,R(z)) & = \frac{3 j_1(k R(z))}{(k R(z))}\,, \label{eq:Window_Fourier}
\end{align}
where $j_1(x)$ is a spherical Bessel function. A straightforward computation gives
\begin{align}
\sigma^2(z,R(z),h) = \int_0^{+\infty} \frac{\dd k k^2}{2\pi^2} b^2(z,k) P(z,k) \t{W}_s^2(k,R(z))\,,\label{eq:sigma_R}
\end{align}
where in practice we apply an infrared and UV cut-off to the integral. The matter power spectrum $P(z,k)$ is defined as
\begin{align}
\langle \delta\e{dm} (z,\bs{k})  \delta\e{dm} (z,\bs{k'})  \rangle=( 2\pi)^3 \delta^{(3)}(\bs{k} - \bs{k'}) P(z,k)\,,
\end{align}
and the bias parameter $b(z,k)$ relates the galaxy and matter density contrast as $\delta_g(z,k) = b(z,k)\cdot \delta\e{dm}(z,k)$. 
Note that we are interested in the variance in the galaxy field and not in the dark matter field, since we assume that GWs are emitted from compact objects, which live inside galaxies. This issue can be circumvented by using the fact that we are looking for a lower bound on $\sigma\e{g}$. Indeed, while knowing the exact value of $\sigma\e{g}$ would be best, introducing any amount of variance in the galaxy field would give a better estimate than homogeneous completion which amounts to setting $\sigma\e{g}=0$. We assume that $b(z,k)\geq 1$, which is a reasonable assumption if we adopt the following argument: in principle, the galaxy bias also depends on the halo mass. However, most stars, which eventually become GW dark siren sources, live in galaxies of $\sim10^{11}M_{\odot}$ (see the bottom panel of Fig.\,12 of \cite{Driver:2022vyh}), for which the bias is expected to be larger than one \cite{Tinker:2010}, at least for large scales. This assumption may break down if the volume of voxels is not large enough compared to the typical size of a galaxy. This introduces a lower redshift cut $z\e{min}$ below which the calculation of $\sigma\e{g}$ cannot be trusted. We assume that the bias is larger than a certain treshold $b\e{th}$ for all relevant galaxies which can host GW sources. We assume that the treshold is $b\e{th} = 1$, although one can easily adapt it, if it turns out to be lower. In the extreme case $b\e{th} \to 0$, which corresponds to no clustering, the method converges to homogeneous completion. 

We use the following symbolic approximation for the non-linear matter power spectrum
\begin{equation}
\begin{aligned}
\ln(P(k, h, z)) & = 4.78 h + \left(0.972 - 1.24 z\right) \left(- 0.152 k + \tanh{\left(1.4 k \right)} + 1.17\right) \\
& \quad - \frac{0.217 k + \tanh{\left(0.727 h - 107.0 k \right)}}{\tanh{\left(0.231 h + 112.0 k \right)}} - \cos{\left(1.37 h - 0.201 k + 0.892 z \right)} \\
&  \quad + \frac{\cos{\left(0.982 h + 0.0826 z \right)}}{0.732 k + 0.132}\,, 
\end{aligned}
\label{eq:Power_Spectrum}
\end{equation}
which is $\sim 10 \%$ accurate in the parameter space identified by the constraints: $0.001 h/{\rm Mpc} < k < 9.4 h/{\rm Mpc}$, $0 < z < 1$, $0.61 < h < 0.73$.\footnote{See appendix~\ref{Sec:symbolic_Pk} for more information on how this expression was obtained.}  Clearly, the dimensionless standard deviation $\sigma(z,R,h)$ given in Eq.\,\eqref{eq:sigma_R} depends on $h$. 
We discuss the cosmology dependence introduced by this in Sec.\,\ref{subsec:Cosmology_Dependence}. We take the value of $b$ which minimizes the amplitude of the standard deviation and define the minimal standard deviation in the galaxy field as
\begin{align}
\sigma\e{g}(z,h) \equiv \hat{\bar{n}}\e{g}(z,h) \cdot \min_{b\geq b\e{th}} \sigma(z,R,h)\,. \label{eq:minimal_sigma_g}
\end{align}
Note that multiplying with the average number density of galaxy, introduces an extra cosmology dependence via the voxel size and the completeness fraction, which enters in Eq.\,\eqref{eq:nhatg}. However, when implementing this method in \texttt{gwcosmo}, mostly the ratio $\sigma\e{g}/\hat{\bar{n}}\e{g}$ is relevant, for which the cosmology dependence arising from the volume of voxels cancels out. We will check the impact of this cosmology dependence in Sec.\,\ref{subsec:Cosmology_Dependence}.

Finally, the integral in Eq.\,\eqref{eq:sigma_R} depends on the UV cut-off $k\e{max}$. For constant $ b(z,k)$, the integrand is a damped oscillating function. The dependence is strongest at the lowest redshift where variance completion applies and where voxels are the smallest. The voxel size already constrained the lowest redshift to be such that $z>z\e{min}=0.015$. At this redshift, we find that $\sigma$ oscillates as a function of $k\e{max}$ with an amplitude of about 10$\%$, which corresponds to the level of precision of our method. At redshift $z=0.1$, these oscillations have an amplitude of about 3$\%$. In the following, we compute $\sigma$ using
\begin{align}
\sigma^2(z,R,h) = \int_{k\e{min}}^{k\e{max}} \frac{\dd k k^2}{2\pi^2} P(h,k,z) \t{W}_s^2(k,R)\,,
\end{align}
with $k\e{min} = 0.001 h/$Mpc, $k\e{max} = 9.4h/$Mpc, $P(k,h,z)$ from Eq.\,\eqref{eq:Power_Spectrum}, $\t{W}_s^2(k,R)$ from Eq.\,\eqref{eq:Window_Fourier} and ignore the weak dependence on the UV cut-off as well as the uncertainty on the power spectrum which are of similar order. 

Eq.\,\eqref{eq:minimal_sigma_g} gives an estimate of the minimal standard deviation in the galaxy number density field that can be used in variance completion. In the following section, we explain how to introduce variance completion in a redshift line of sight prior from \texttt{gwcosmo}.

\subsection{Implementation in \texttt{gwcosmo}}\label{subsec:gwcosmo}

Given the level of sophistication of \texttt{gwcosmo}, it turns out to be more convenient to condense the information of variance information into a ratio function $R(z,\bs{\hat{n}})$, which multiplies the out-of-catalogue contribution to the redshift line of sight prior of \texttt{gwcosmo} rather than to apply Eq.\eqref{eq:LOS}. This allows to easily use other features of the line of sight redshift prior such as luminosity weighting. Schematically, we write that the probability density of presence of a galaxy at redshift $z$ in the pixel corresponding to direction $\bs{\hat{n}}$ is given by
\begin{align}
p(z,\bs{\hat{n}}) = p\e{c}(z,\bs{\hat{n}}) + R(z,\bs{\hat{n}}) p\e{m}(z,\bs{\hat{n}})\,.\label{eq:LOS_pdf} 
\end{align}
This corresponds to Eq.\,(2.22) of \cite{Gray:2023wgj} where the first term indicates the in-catalogue contribution, while the second term is proportional to the homogeneous out-of-catalogue contribution $p\e{m}(z,\bs{\hat{n}})$. The ratio function $R(z,\bs{\hat{n}})$, introduced here, returns $1$ whenever variance completion is uninformative and otherwise redistributes power from voxels which are more likely to be underdense to voxels which are more likely to be overdense. The ratio function is computed as follows
\begin{align}
R(z,\bs{\hat{n}}) \equiv \frac{\hat{n}\e{m;var}(z,\bs{\hat{n}})}{\hat{n}\e{m;hom}(z,\bs{\hat{n}})}\,.\label{eq:Ratio_function}
\end{align}
This has the benefit to cancel what ever modeling we used for $\hat{n}\e{m;hom}$ since $\hat{n}\e{m;var}$ averages to $\hat{n}\e{m;hom}$. In this context, variance completion therefore boils down to estimate this ratio function in each pixel and redshift. 

We apply variance completion to B-band galaxies of GLADE+ and luminosity weight them as per Eq.\,\eqref{eq:M_to_L}. The high density to higher redshifts of the B-band with respect to the K-band allows us to successfully apply variance completion and to estimate the ratio function via Eq.\,\eqref{eq:Ratio_function}. Indeed, there are about 20M galaxies galaxies in the B-band against, 1M in the K-band. We then combine this ratio function with \texttt{gwcosmo} lines of sight extracted for the K-band. Note that while this is not strictly necessary, this allows to combine information from different bands. We work with bin sizes of $\delta z = 0.002$ between redshift $z\e{min} =0.015$ and a maximum redshift $z\e{max}(\bs{\hat{n}}) \in [0.07,0.20]$ which depends on direction. The maximal redshift is set such that the completeness in the B-band is above $10\%$. We get a nontrivial value for the ratio function for the voxels in the interval $z\in [z\e{min},z\e{max}(\bs{\hat{n}})]$ and we set it to $1$ everywhere else. 

\begin{figure}[ht]
\centering
\includegraphics[width=1.0\textwidth]{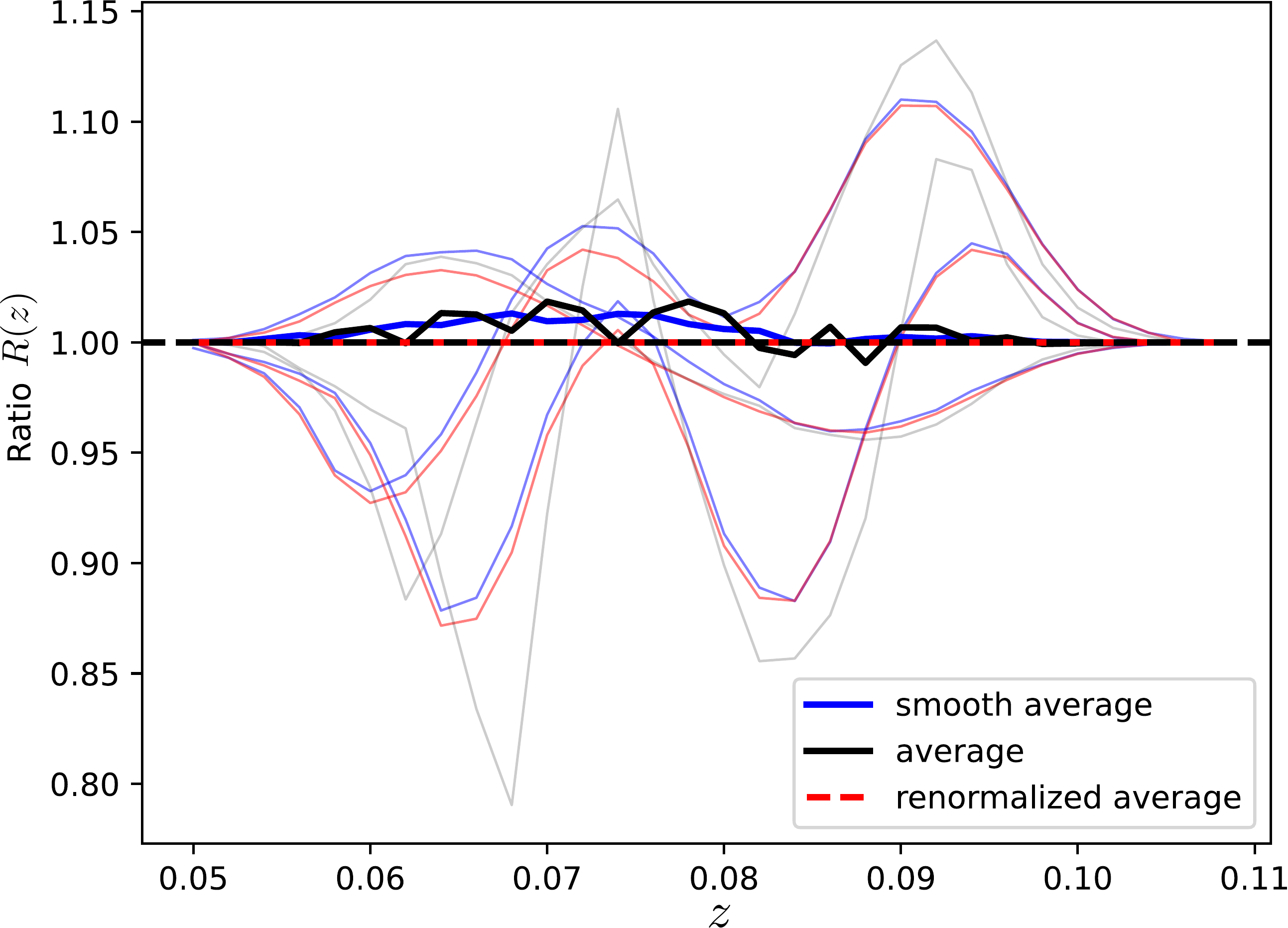}
\caption{We plot 3 lines of sight at different levels of postprocessing. In black, we show the outcome of variance completion, after filling the gaps and removing values which are below the threshold of $\bar{n}\e{m}/10$. This explains why the average over the region $\S$ (thick black) differs from $1$. In blue, we show the same curves after Savitzky-Golay filtering. In red, we show the same curves after renormalizing them such that the average of $R(z)$ over the 1024 lines of sight of this region average to 1 at each redshift $z$. The thick curves show the average over the region $\S$ at each stage of postprocessing.}  
\label{fig:Ratio_example_processing}
\end{figure}

\begin{figure}[ht]
\centering
\includegraphics[width=1.0\textwidth]{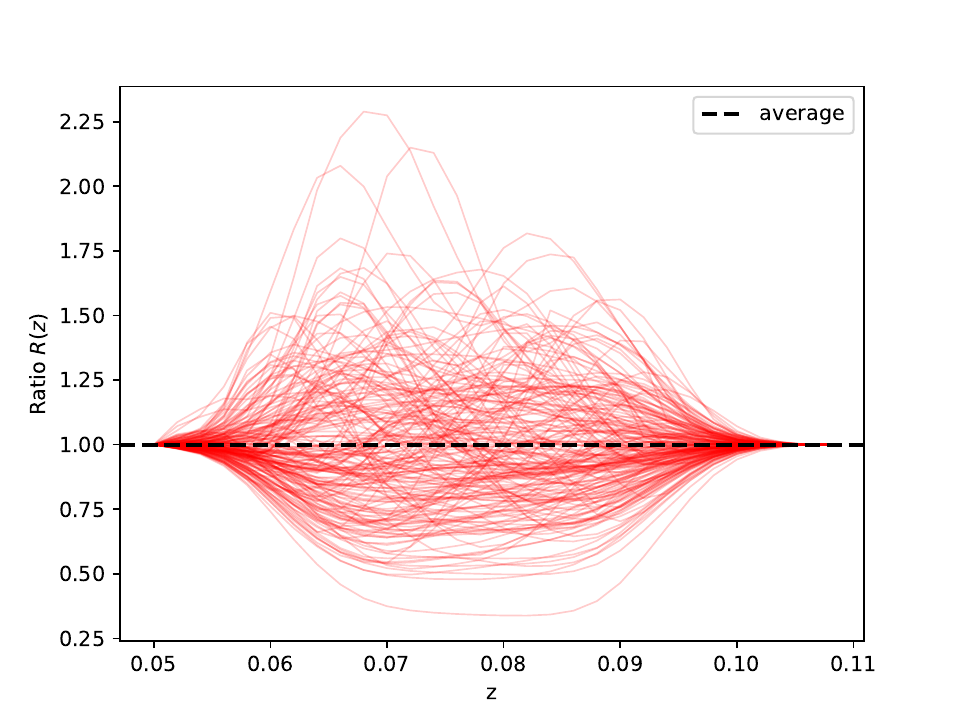}
\caption{We plot 200 ratio functions $R(z,\bs{\hat{n}})$ for 200 lines of sight in a same region $\S$. Some lines of sights have less galaxies over the relevant redshift range, implying that variance completion sets them to a value $R<1$.
However, they are expected to be as complete as other lines of sight, based on their magnitude threshold, which is probably dominated by galaxies seen at lower or (higher) redshift. The peaks at $R(z)\sim 2$ denote that these lines of sight are likely to host an over-dense region such as a cluster or a filament at this redshift. This class of lines of sight has a maximum redshift $z\e{max}(\bs{\hat{n}})=0.105$, where the completeness in the B-band drops below $10\%$. }  
\label{fig:Ratio_example_200}
\end{figure}

When the algorithm of variance completion does not converge on 1 or 2 consecutive redshift slices, we interpolate between those (This happens for less than 10\% of voxels). If it does not converge on more consecutive slices, we interpolate smoothly with homogeneous completion. The outcomes of $\hat{n}^i\e{m;var}$ which are below a threshold (fixed to $n\e{t} \equiv \hat{n}\e{m;hom}/10$) are filled with a minimum threshold value of $n\e{t}$ (This happens for less than $0.1\%$ of the voxels to which we applied the algorithm). We then use a \textit{Savitzky-Golay} filter to smooth the curves. Because of the postprocessing and numerical inaccuracies in variance completion, averaging the lines of sight ratio functions in a class $\S$ at fixed redshift does not necessarily average to unity. We correct for this by renormalizing each line of sight to ensure that the ratio functions of a class $\S$ at fixed redshift average to one. We show the process from the results of variance completion for the ratio function to a smooth ratio function that we use in \texttt{gwcosmo} in Fig.\,\ref{fig:Ratio_example_processing}. We show $200$ lines of sight ratios in Fig.\,\ref{fig:Ratio_example_200}. 

We do not expect the ratio function to affect the normalization of the entire line of sight redshift prior $p(z,\bs{\hat{n}})$. This is because of the following identity
\begin{align}
\int_{\mathbb{S}_2} \dd \Omega \, R(z,\bs{\hat{n}}) \cdot p\e{m}(z,\bs{\hat{n}}) & = \sum_{i}\int_{\S_i} \dd \Omega \, R(z,\bs{\hat{n}}) \cdot p\e{m}(z,\bs{\hat{n}}) \\
& \simeq \sum_{i}  p\e{m}(z,\S_i) \underbrace{\int_{\S_i} \dd \Omega\, R(z,\bs{\hat{n}})}_{=1} = \int_{\mathbb{S}_2} \dd \Omega \, p\e{m}(z,\bs{\hat{n}})\,, 
\end{align}
where the last approximate equality holds because the out-of-catalogue contributions have similar completeness levels in a fixed region $\S_i$. This implies that
\begin{align}
 \int_0^{z\e{m}}  \int_{\mathbb{S}_2} \dd z \dd \Omega\, p(z,\bs{\hat{n}}) \simeq \int_0^{z\e{m}}  \int_{\mathbb{S}_2}  \dd z \dd \Omega \l( p\e{c}(z,\bs{\hat{n}})+ p\e{m}(z,\bs{\hat{n}})\r)
\end{align}
such that the normalization is approximately preserved.

The ratio function introduced in Eq.\,\eqref{eq:LOS_pdf} and \eqref{eq:Ratio_function} allows to separate completely the determination of variance completion from the traditional line of sight redshift prior computation. This permits to easily implement variance completion in a dark siren method based on the line of sight redshift prior such as \texttt{gwcosmo} or \texttt{icarogw}.

\subsection{Cosmology dependence of the line of sight redshift priors}\label{subsec:Cosmology_Dependence}

Ultimately, we aim to use these lines of sight redshift priors to estimate the Hubble constant $H_0$. Since we have used $H_0$ to determine $\sigma\e{g}$, the completeness fraction and the voxel volumes, it is legitimate to ask what is the impact of the choice of $H_0$ on the lines of sight. For our purpose, we stick to the dependence on the reduced Hubble constant $h$, defined as $H_0 \equiv 100 h$ km s$^{-1}$ Mpc$^{-1}$, leaving any dependence on $\Omega\e{m0}$ and the dark energy equation of state to future work. In the end, we implement the method via the ratio function $R(z,\bs{\hat{n}})$, given in Eq.\,\eqref{eq:Ratio_function}. Because of the many steps required to reach an estimation of $\hat{n}\e{m;var}$ and $\hat{n}\e{m;hom}$ which enter in its definition, it is not entirely obvious what is the cosmology dependence of the ratio function. For instance, the standard deviation $\sigma(z,R,h)$ depends on $h$ in two ways. First, through the power spectrum $P(k,h,z)$ which appears in \eqref{eq:sigma_R}. Second, it depends on $h$ implicitly through the sphere equivalent radius $R$ of the voxels, which appears in the same expression. This one depends on cosmology because it is extracted from the comoving volume of voxels of fixed angular size and fixed redshift bin. Another non-obvious example is the dependence of $\hat{n}\e{m;hom}$ on the completeness fraction $\hat{f}_\S$ which is made through the apparent magnitude threshold map, converted to a luminosity threshold. This last step depends on cosmology via Eq.\,\eqref{eq:m_and_z_to_M}-\eqref{eq:M_to_L}. Finally, the estimates $\hat{n}\e{m;var}^i$ are obtained by solving numerically and iteratively a linear system, which depends nonlinearly on $\sigma\e{g}$. This makes the cosmology dependence of the ratio function intractable analytically; instead we will assess it numerically. 

\begin{figure}[ht]
\centering
\includegraphics[width=1.0\textwidth]{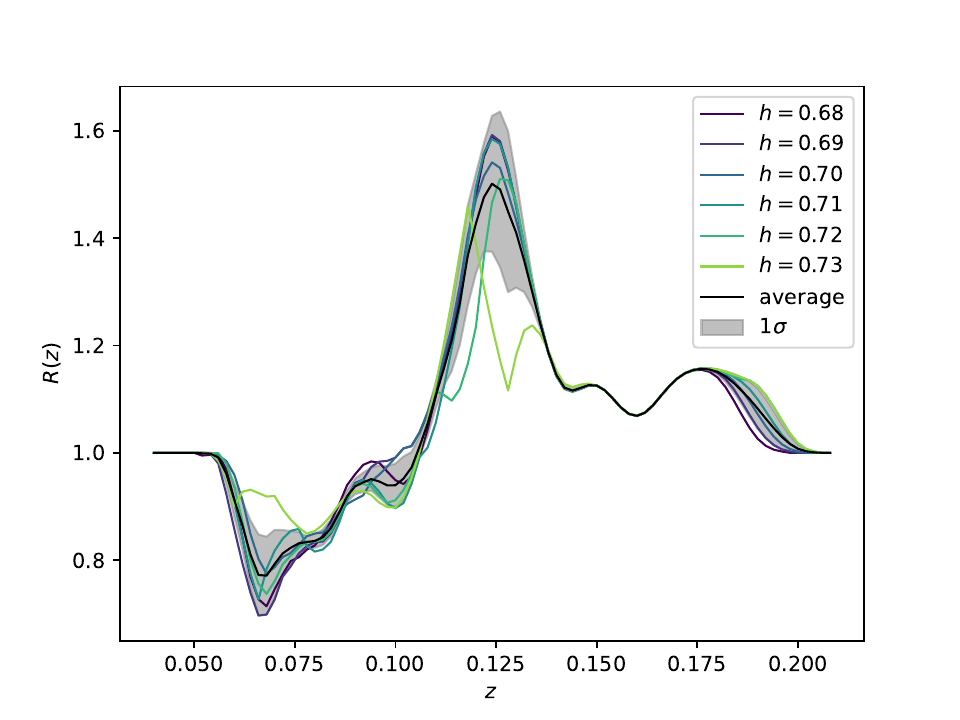}
\caption{We plot an example of ratio function for different values of $h$. The grey shaded area indicates the 1 sigma contour around the average, which is in black. First, one can recognize the dominant features from one value of $h$ to another. Second, the relative amplitude of the peaks is not preserved from one value of $h$ to the other. There are several reasons for this. One reason is that variance completion is anyways approximate. Indeed, the variance completion algorithm stops when the total number of missing galaxies introduced matches the expectation from homogeneous completion within $1\%$. A further reason is the postprocessing (gap filling and smoothing, described in Sec.\,\ref{subsec:gwcosmo}) of the lines of sight, which easily modifies the shape of the lines of sight to an extent which is larger than what can be seen here. The variance completion also applies to a higher maximum redshift for higher $h$. The reason for this is that the completeness is cosmology dependent and we stop applying variance completion beyond a lower threshold on completeness, which is cosmology independent.}  
\label{fig:Cosmology_dependent_R}
\end{figure}

Since the out-of-catalogue part is modulated by the ratio function, it is sufficient for our purpose to study the $h$ dependence of the ratio function, which contains in a single function of redshift and pixel, all the practical $h$ dependence of the variance completion method. To understand qualitatively, the behavior of the ratio function under a change in $h$, we plot 1 line of sight for different values of $h \in \mathcal{H}\equiv\{ 0.67,0.68,0.69, 0.70,0.71,0.72,0.73\}$ in Fig.\,\ref{fig:Cosmology_dependent_R}.
The ratio function of this particular line of sight represents well the fact that we get different results for different values of $h$. However, the overall structure of the features is preserved. Given that the postprocessing seems to induce just as much change in the ratio function as a change in $h$ and the fact that the power spectrum is precise to about $10\%$, we conclude that the $h$ dependence is subdominant and acceptably negligible to within our current level of precision. To remove any remaining dependence on $h$, we marginalize over $h \in \mathcal{H}$. We reconstruct the line of sight using the average ratio function
\begin{align}
R(z,\bs{\hat{n}}) = \langle R(z,\bs{\hat{n}},h) \rangle_h\,.
\end{align}
Therefore, strong features survive this average if they appear independently of $h$. Their amplitude is moderated by that average. We plot an example line of sight for these values of $h$ with the corresponding average and $1\sigma$ contours in Fig.\,\ref{fig:Cosmology_dependent_R}.
It should be noted that in applying this average, we have used a much narrower prior on $h$ than \texttt{gwcosmo} does. Indeed, the later assumes the conservative prior $h\in[0.2,1.4]$. Given, that this information only feeds in a fraction of the galaxy data, for which there is evidence that $h$ is consistent with the prior $\mathcal{H}$ \cite{DESI:2024mwx}, we do not expect this to affect significantly our conclusions. Another possibility is for each pixel and at each redshift, to take the minimum on $h$ of $R(h,z)$. In this way, one ensures to take the most conservative prior on structure. Alternatively, one may divide the ratio excess from unity by a number larger than 1 to suppress the amplitude of the ratio function. This would also result in a more conservative estimate on structure. Note that this is the analogue of assuming a smaller minimum bias treshold for the calculation of $\sigma\e{g}$ but applied to the results of variance completion instead of the input.

Here we have discussed the cosmology dependence of the ratio function and marginalized over the remaining dependence. This allows us to construct a line of sight redshift prior based on variance completion and combine with GW data to extract the Hubble constant, as discussed in the next sections.

\section{Results}
In this section, we present the redshift line of sight priors determined from variance completion applied to the GLADE+ catalogue. We then perform a dark siren analysis of GW190814, a well localized event, which has good catalogue support. Finally, we apply the method to the events up to the third observing run, restricting to events with SNR$>11$.

\subsection{Updated line of sight redshift priors}
\begin{figure}[ht]
\centering
\includegraphics[width=1.0\textwidth]{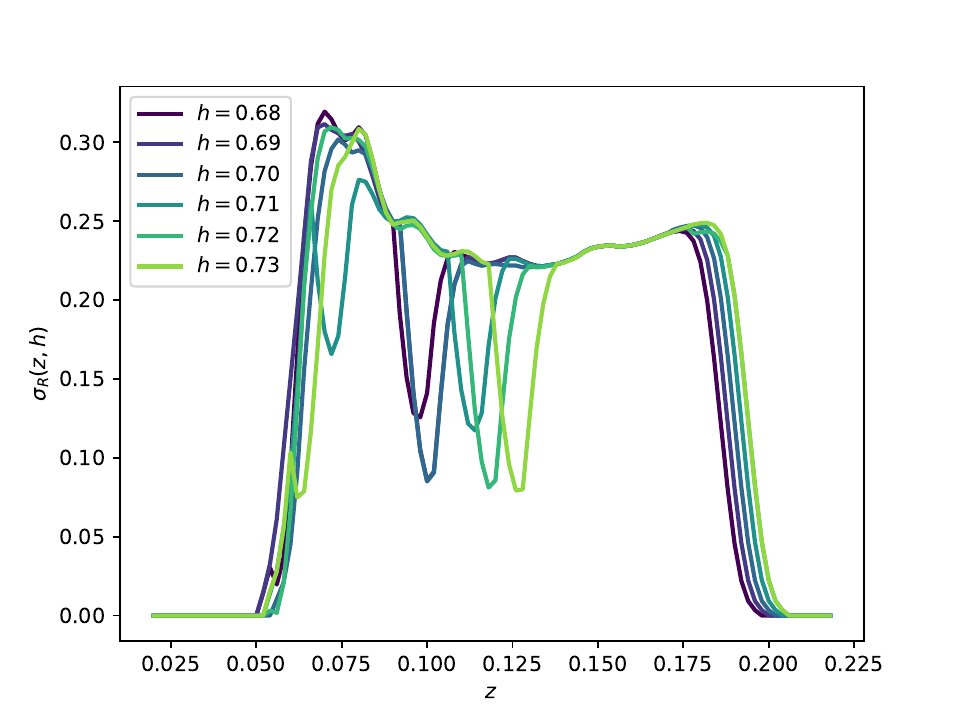}
\caption{Standard deviation $\sigma_R(z,h)$ of the ratio function $R(z,h)$ as a function of redshift $z$ for different values of $h$. At fixed redshift $z$, there exists a value of $h$ which minimizes $\sigma_R(z,h)$, giving a prospective $h$ which minimizes the amplitude of structure at that redshift. However, that value changes with redshift, making the value of $h$ which minimizes structure difficult to identify unequivocally. For certain values of $h$ and at certain redshifts, the amplitude of the dispersion is lower than for the other values of $h$. That can happen if variance completion converges to a solution with lower dispersion. }  
\label{fig:sigmaR(z)}
\end{figure}

We apply variance completion with $n\e{side} =32$ and $\delta z = 0.002$ between redshift $z\e{min} =0.015$ and a maximum redshift $z\e{max}(\bs{\hat{n}})\in [0.07,0.20]$, which is fixed such that the B-band completeness remains above $10\%$. In this interval, the typical variance enforced by variance completion on the out-of-catalogue contribution is such that the standard deviation in the ratio function $R(z)$ is about $\sigma_R \lesssim 0.3$, as may be seen in Fig.\,\ref{fig:sigmaR(z)}. Outside these domains, we smoothly interpolate with homogeneous completion. The complete methodology is explained in details in Sec.\,\ref{sec:Methodology}.

\begin{figure}[ht]
\centering
\includegraphics[width=1.0\textwidth]{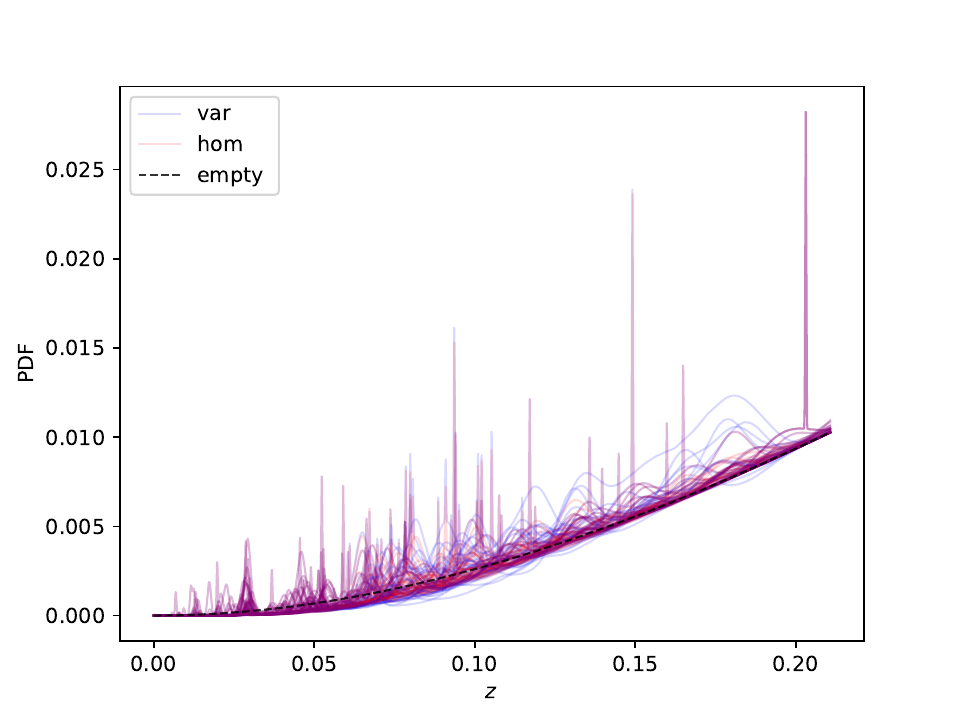}
\caption{We plot 50 lines of sight for variance completion and homogeneous completion. At low redshift, when the completeness of GLADE+ is relatively high, they overlap well. It is only at higher redshift that variance completion takes a little bit more risk by assessing which of the lines of sight are more likely to be void or clusters depending on their observed number counts and on prior knowledge of large scale structure. Beyond a certain redshift, variance completion smoothly interpolates with homogeneous completion.}  
\label{fig:LOS50}
\end{figure}

In Fig.\,\ref{fig:LOS50}, we present 50 lines of sight for variance completion versus homogeneous completion using luminosity weighting in both cases. One clearly sees that they overlap well at lower redshift and start to differ when the completeness drops, at higher redshift. Combined, the lines of sight average to the empty catalogue prior, which is also plotted there. Variance completion redistributes probabilities for the out-of-catalogue contribution to the redshift LOS prior, from voxels which seem to be emptier to those which are expected to contain more galaxies, at fixed redshift. Finally, between redshift $0.15$ and $0.20$, one can see some lines of light which are interpolated between variance completion and homogeneous completion. In the next section, we apply these line of sight redshift priors to the event GW190814, which is a well localized GW event.

\begin{figure}[ht]
\centering
\includegraphics[width=1.0\textwidth]{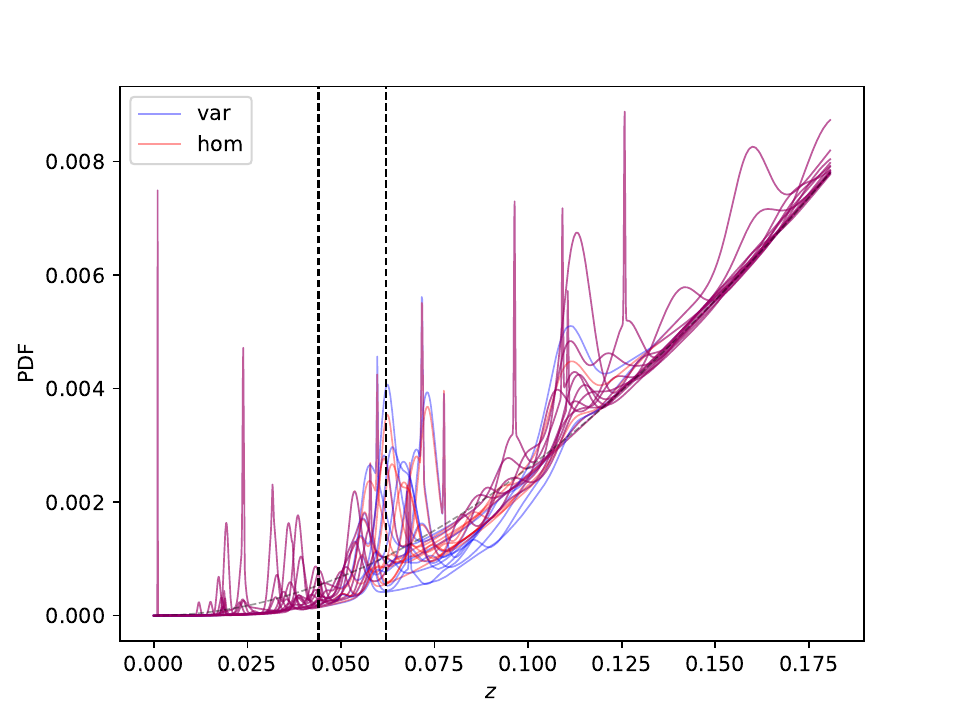}
\caption{Here, we plot the 13 lines of sight corresponding to the $99.9\%$ confidence sky localization of GW190814 for homogeneous completion (in red) and variance completion (in blue) both with in-catalogue galaxies weigthed by their K-band luminosity. The vertical dashed lines shows the $\pm 1\sigma$ contour for the expectation on the redshift of the GW source. The dashed grey line shows the empty catalogue line of sight redshift prior. } \label{fig:LOS_var_GW190814}
\end{figure}

\begin{figure}[ht]
\centering
\includegraphics[width=1.0\textwidth]{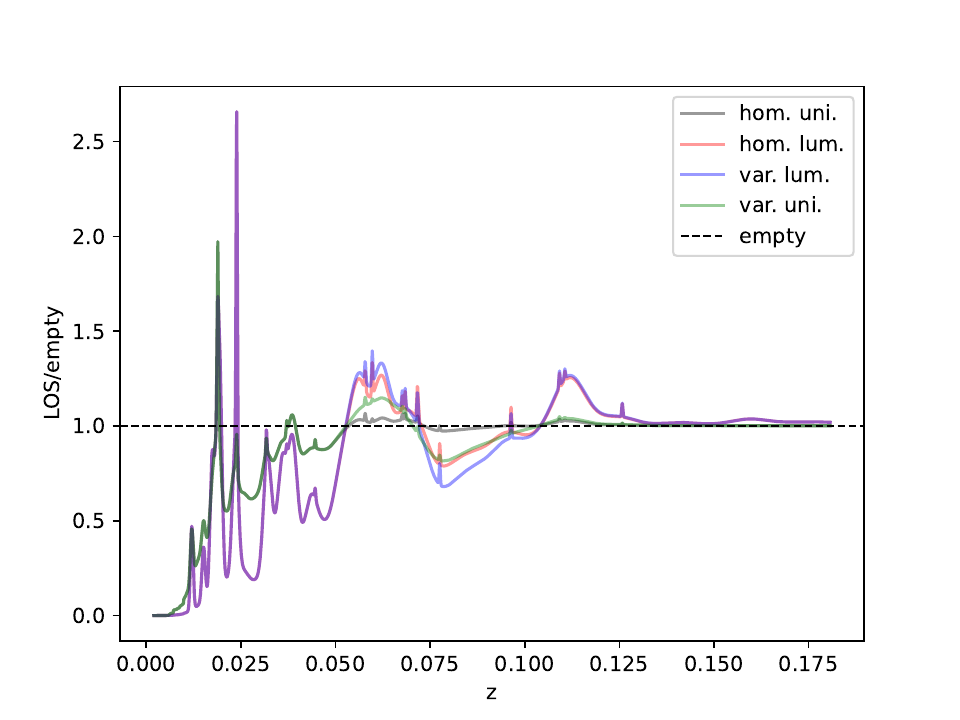}
\caption{We plot the ratio of the combined line of sight redshift prior for the 13 pixels corresponding to the $99.9\%$ confidence sky localization for different galaxy weightings and completion. We plot homogeneous completion with uniform (black) and luminosity (red) weighting of in-catalogue galaxies. We also plot variance completion with luminosity weighting (in blue) and with uniform weighting in green. The change is less spectacular than for individual lines of sight. This should not be surprising since we are averaging 13 lines of sight. However, for both uniform and luminosity weighting, variance completion is more informative than homogeneous completion, in the sense that the lines of sight differ more from the empty catalogue. On the right of the plot, the catalogue converges to an empty one which explains why the ratio asymptotes to unity.}  
\label{fig:LOS_var_over_empty}
\end{figure}

\subsection{Dark siren analysis of the Hubble constant for GW190814}
\begin{figure}[ht]
\centering
\includegraphics[width=1.0\textwidth]{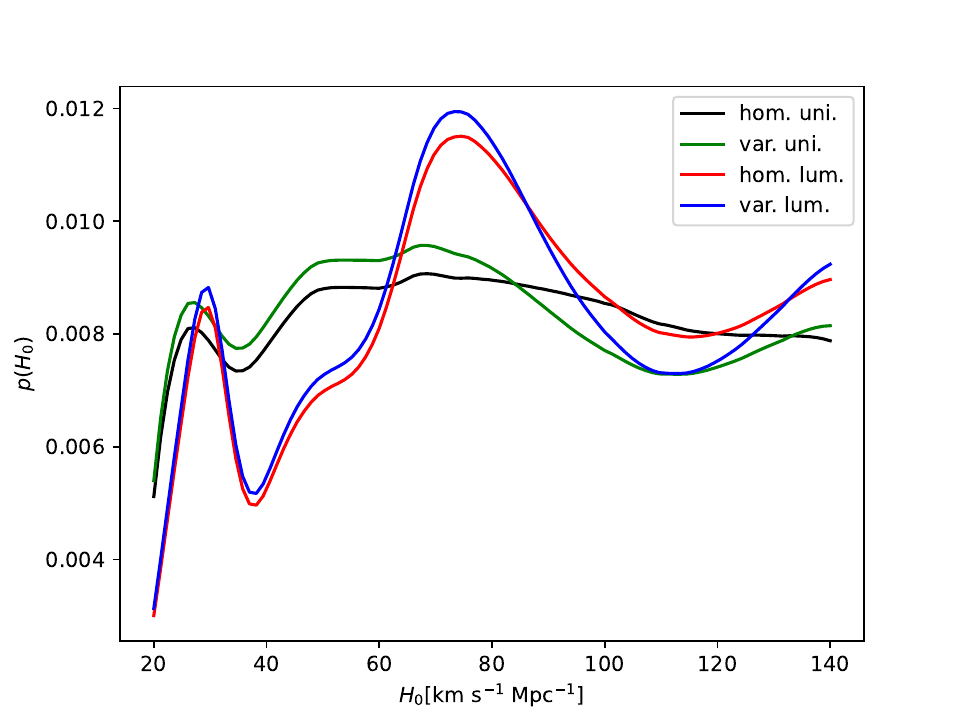}
\caption{Here, we present the posterior for the Hubble constant $H_0$ for three different line of sight prescriptions for GW data coming from GW190814. The median and $68.3\%$ confidence intervals for uniform weighting of in-catalogue galaxies, homogeneous and variance completion give $H_0 =79.373^{+40.560}_{-39.054}, 76.510^{+43.070}_{-37.246}$ km s$^{-1}$ Mpc$^{-1}$, respectively. For luminosity weighting of in-catalogue galaxies, homogeneous and variance completion give $H_0 = 82.122^{+39.153}_{-36.991}$ and $80.351^{+40.770}_{-36.213}$ km s$^{-1}$ Mpc$^{-1}$, respectively. In both the uniform and luminosity weighting cases, variance completion suppresses power at relatively high values (between $90-130$) of $H_0$ with respect to homogeneous completion. The curves have been smoothed to remove unphysical features coming from the finite number of $2\cdot 10^6$ injections. } 
\label{fig:posterior_H0}
\end{figure}

With our brand new LOS redshift prior, we analyze the GW event GW190814, which has fairly good catalog support and excellent sky localization ($90 \%$ confidence regions covers about 19 deg$^2$). This event has also been analyzed for astrophysics and tests of gravity \cite{LIGOScientific:2020zkf}. We present the 13 lines of sight which correspond to the $99.9\%$ sky localization region of GW190814 in Fig.\,\ref{fig:LOS_var_GW190814}. We also plot the $68\%$ confidence interval for the expected redshift $z$ of GW190814, given the luminosity distance measurement and assuming a Planck prior on $H_0$. This region overlaps well with a place where the impact of variance completion begins to be noticeable by eye on individual lines of sight. For single lines of sight, the change can be relatively important. However, combining as few as 13 lines of sight with equal weights, averages out some of the structure brought in by variance completion, as may be seen in Fig.\,\ref{fig:LOS_var_over_empty}. Indeed, assuming for simplicity that the over- and under-densities of these $N$ lines of sight are gaussian distributed, we expect the standard deviation at a given redshift to be of order $\sigma = \sqrt{\sigma_R^2(z)/N}$. For $\sigma_R(z)\sim 0.3$ as may be seen in Fig.\,\ref{fig:sigmaR(z)} and $N = 13$, this gives $\sigma \simeq 0.08$. This corresponds roughly to the amplitude of the difference between the two curves in Fig.\,\ref{fig:LOS_var_over_empty}. It should be clear that if the sky localization covers more pixels, then the information brought in by variance completion averages out even more, although weighting each pixel by their respective probability to host the GW may mitigate this effect. In Fig.\,\ref{fig:posterior_H0}, we plot a Hubble constant posterior using the uniform and luminosity weighting with homogeneous completion versus luminosity weighting with variance completion. It turns out that variance completion gives a posterior distribution which is marginally less compact for this particular event. Uniform or luminosity weighting with homogeneous completion and luminosity weighting with variance completion give $H_0 =79.373^{+40.560}_{-{39.054}}, 82.122^{39.153}_{-36.991}, 80.353^{+ 40.770}_{-36.213}$ km s$^{-1}$ Mpc$^{-1}$ respectively. They are all consistent with each other and with the results of \cite{LIGOScientific:2020zkf,Vasylyev:2020hgb}, which use \texttt{gwcosmo} which relies on homogeneous completion. They are also consistent with the analysis of \cite{DES:2020nay}, which use the Dark Energy Survey, which is complete at the relevant redshift range and as such, does not require any completion technique. In the next section, we apply variance completion to GW events of up to the third observing run of LVK.

\subsection{Application to GWTC-3}
\begin{figure}[ht]
\centering
\includegraphics[width=1.0\textwidth]{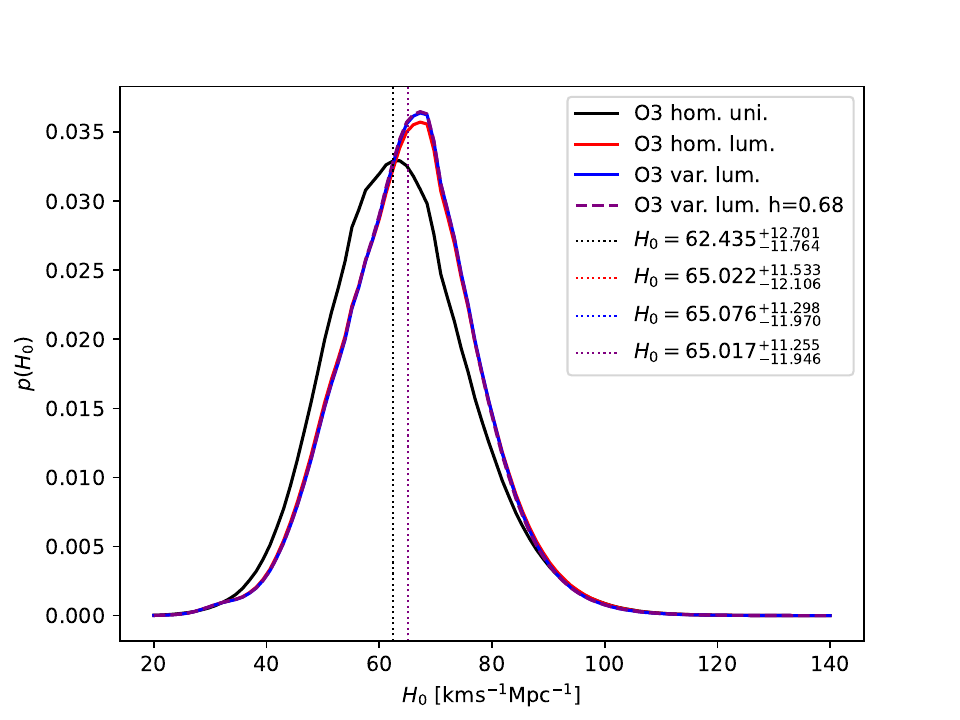}
\caption{We present the posterior on $H_0$ for 43 events of O3 for homogeneous completion with uniform weighting (O3 uni), homogeneous completion with luminosity weighting(O3 lum), variance completion with luminosity weighting (O3 var) and variance completion using $h=0.68$ with luminosity weighting. These last two curves mentioned, sit almost on top of each other. The reported values of $H_0$ are in units of km s$^{-1}$ Mpc$^{-1}$ and correspond to the median and 68.3$\%$ confidence intervals. } 
\label{fig:O3_posterior_H0}
\end{figure}

We apply the line of sight priors generated with variance completion to the third observing run of LVK using 46 events which have SNR$>11$, including binary black hole and neutron star black hole events. Compared to GW190814, analyzed in the previous section, those events span very different levels of uncertainty in the localization solid angles $\Delta \Omega$. The best localized events have $\Delta\Omega = \mathcal{O}(10)$ deg$^2$, while the worst localized event has $\Delta\Omega = 14,000$ deg$^2$. We obtain the posterior on $H_0$ presented in Fig.\,\ref{fig:O3_posterior_H0} for homogeneous and variance completion, using luminosity weighting in both cases. We also plot the result of homogeneous completion with uniform weighting of galaxies. The measurement of $H_0$ changes from $H_0 =(65.022^{+11.533}_{-12.106}) $ km s$^{-1}$ Mpc$^{-1}$ for homogeneous completion to $H_0 = (65.076^{+11.298}_{-11.970})$ km s$^{-1}$ Mpc$^{-1}$ for variance completion. We also show the result for variance completion when $h=0.68$ is assumed to generate the ratio function for variance completion. The three measurements are consistent with each other and with the results of \cite{LIGOScientific:2021aug}, which use \texttt{gwcosmo} with homogeneous completion. They are also consistent with the results of \cite{Palmese:2021mjm} which use the DESI legacy survey and deal with incompleteness with a data driven approach which avoids having to model the luminosity function. Our results are also consistent with those of \cite{Alfradique:2023giv}, which use dark siren events for which the galaxy catalogue DELVE is complete over the corresponding GW sky areas. Finally, the authors of \cite{Bom:2024afj} focus on the DESI legacy survey and DELVE and focus on GW events for which they expect their galaxy catalogues to be complete. 

The large scale structure prior changes slightly the median value of $H_0$ and makes the error bars slightly smaller. This is what we expect from adding in information from the large scale structure for the out-of-catalogue galaxies. The reason why variance completion does not perform significantly better than homogeneous completion is a reflection of what happens for the best localized event GW190814. As we argued in the previous section on this event, averaging the line of sight redshift prior over as few as 13 lines of sight, already washes significantly the effect of variance completion. For the worse localized event, implementing variance completion over many more lines of sight (about $4,170$ for $\Delta\Omega = 14,000$ deg$^2$ ), the out-of-catalogue contribution simply averages out to the empty catalogue. This means that only a handful of events with good sky localization ($\Delta \Omega = \mathcal{O}(10)$ deg$^2$) contribute to the improvement of the line of sight redshift prior.

\subsection{Discussion} 
The goal of this work is to maximize the potential constraining power of dark siren analyses, by including large scale structure knowledge in the method. Implementing variance completion is a step towards this goal, but to achieve this we had to overcome several challenges. 

One such difficulty is the resolution in pixel and in redshift space. Indeed, we had to deal with \texttt{healpix} pixels and redshift bins, with voxels which change geometry with redshift. The ideal resolution for variance completion to work is a compromise between two extremes. If the resolution is too low, then the variance in the number counts on large volumes drops to zero, making the method converge to homogeneous completion. If the resolution is too high, the variance in the number count can be significant, potentially maximizing the outcome of variance completion, but four other limitations appear. First, the number of voxels required to approximate a homogeneous patch of the Universe ($\sim 10^6$ Mpc$^3$) on which the method was designed to work becomes too large ($> 1000$), making the method computationally heavy. Second, at some point the smoothed voxalated galaxy number densities do not accurately represent the galaxy field distribution anymore. Third, we need an input on the variance in the galaxy field, which we could bound from below, by assuming that the bias satisfies $b(z,\bs{k})\geq 1$. We cannot apply this bound anymore if we lower the resolution down to arbitrary small scales, in which case, the concept of bias does not make sense anymore. Finally, the lowest resolution that we can apply the formalism to, depends on what scales we manage to model the nonlinear power spectrum. In our case, $k\e{max} =9.4 h/{\rm Mpc}$, which corresponds to a wavelength of approximately $1$ Mpc. The fact that the redshift range over which the GLADE+ catalogue is sufficiently empty, allowing variance completion to enhance the dark siren method, and not too empty, to allow the method to work, has resulted in constraints on the geometry. Indeed, we fixed the \texttt{healpix} resolution to $n\e{side} = 32$ and the redshift bin to $\delta z=0.002$. The bin size could well vary with redshift and the healpix resolution changed at various redshift checkpoints, but we leave these possibilities to future work.

One obvious difficulty of this method is the modeling required for $\bar{n}\e{g}$ and $\sigma\e{g}$. The former, which is estimated from data, eventually drops out of the method when implemented in \texttt{gwcosmo}, since it cancels between the numerator and the denominator of the ratio function $R(z,\bs{\hat{n}})$ given in Eq.\,\eqref{eq:Ratio_function}. The variance $\sigma\e{g}^2$ on the other hand is estimated from the power spectrum, which has to be modelled. This required sophisticated techniques presented in the appendix \ref{Sec:symbolic_Pk}. We also simplified the analysis by neglecting geometric effects and by assuming that it is sufficient to consider the voxels as sphere-equivalent. Even if this approximation turns out to be insufficient, the difference between the sphere-equivalent variance $\sigma\e{g}^2$ and the one calculated from the full geometry of a voxel can be absorbed into an assumption of a lower bias prescription than $b(z,\bs{k})>1$. Finally, we assumed that the bias between galaxies and dark matter is larger than unity, which is expected for large scales and which may break down for smaller scales, although we do not expect to enter this regime, given our minimal voxel size.

Redshift uncertainties can be incorporated in the method by integrating the redshift posterior of galaxies over the voxel redshift range. For simplicity, we have limited ourselves to the assumption of Gaussian uncertainties, although this may be inaccurate, especially for photometric redshifts. On top of this, below a certain magnitude threshold, photometric redshifts can be biased measurements of redshift, which is a complication beyond the scope of this work.

Between the raw outcome of variance completion and the smooth curves of the ratio function, we had to apply some postprocessing, to fill gaps, smooth the curves and renormalize the ratio function such that it averages to unity among a class of voxels. We also averaged the ratio function over a few values of $h$. These procedures affect the shape of the lines of sight. As such they may affect the outcome on $H_0$, which is however, difficult to quantify. As the change in the posterior on $H_0$ for 46 events remains small, the effect is negligible for now, but may not remain so in the future.

With the advent of new generations of GW detectors and galaxy surveys, galaxy catalogues will be more complete to higher redshift with better redshift uncertainties and the GW localization volumes smaller. As both of these improve, the method will be relevant to more GW events and should contribute to lower the error bars on the Hubble constant. Meanwhile, the method can always be implemented in parallel to check that homogeneous completion is consistent with the expectation from large scale structure.

While the change may not be far-reaching for now, there may be GW events in the future for which localization volume corresponds only to a few lines of sight, for which variance completion predicts a $2\sigma$ over/under-density in the region of interest. This would have a significant impact on the $H_0$ posterior. Likewise, the identification of multiple images of a strongly lensed GW may allow to constrain their sky localization to a sub-arcsecond level \cite{Hannuksela:2020xor}. This subpixel localization would also result in a higher impact of the out-of-catalogue contribution from variance completion.

\section{Conclusion}

In this work, we have developed variance completion to make it applicable to real galaxy data. After a brief reminder of variance completion, we detailed how to obtain two important ingredients for the method, which are the average number density of galaxies together with its standard deviation. We argued that the former drops out eventually, when implemented in \texttt{gwcosmo} and discussed the cosmology dependence of the latter. We presented the classes of pixels, with similar estimated completeness fractions over which the variance completion algorithm applies. We introduced a ratio function which contains the information on whether we expect a certain voxel to rather be an over or underdensity and with what amplitude. This ratio function can easily be incorporated in the \texttt{gwcosmo} formalism by multiplying the out-of-catalogue homogeneous in comoving volume contribution. We applied some postprocessing to fill gaps, smooth the curves and renormalize them. We showed how this ratio function is not expected to change the overall normalization of the redshift line of sight prior.

We then presented updated lines of sight and applied the method to GW190814, which is a well localized event. The updated lines of sight compare easily with the ones for homogeneous completion and are slightly more informative on some of them, increasing the amplitude of large scale under/over-densities. We then applied the method to extract the Hubble constant and found consistent results whether we applied homogeneous completion or variance completion. The median value shifted a little bit and the error bars as well. For simplicity, it seems preferable to use homogeneous completion for now, even though there is a marginal gain in constraining power from variance completion. This picture may change for GW events that are localized to about $3$ [deg$^2$], which corresponds to one \texttt{healpix} pixel of $n\e{side}=32$. In this case, we can expect variance completion to improve significantly the posterior on the Hubble constant.

In the near future, a wealth of galaxy data is going to be collected from the Large Synoptic Survey Telescope (LSST) \cite{LSST:2008ijt}, Euclid \cite{Euclid:2024yrr} and the Dark Energy Spectroscopic Instrument (DESI) \cite{DESI:2024mwx}. For relatively low redshift regions, the incompleteness may no longer be a problem, although this always depends on the mask. Nonetheless, there will still be redshift regions for which the galaxy catalogues are incomplete. In this work, we have shown in practice how to apply variance completion to a real survey and take into account knowledge of the large scale structure, as a prior for a dark siren analysis.

\acknowledgments

C.D. and T.B. are supported by ERC Starting Grant SHADE (grant no.\,StG 949572). B.F. and T. B. are further supported by a Royal Society University Research Fellowship (grant no.URF\textbackslash R\textbackslash 231006). We thank Rachel Gray, Matthieu Schaller, Konstantin Leyde, Anson Chen for useful discussions. C.D.\,thanks SPCS of QMUL for hospitality. Some of the results in this paper have been derived using the healpy and HEALPix package \cite{Zonca2019,healpix:2005}.

\appendix
\section{Approximate matter power spectrum}
\label{Sec:symbolic_Pk}
The non-linear matter power spectrum is a non-closed form function that requires expensive cosmological simulations to be accurately predicted for each cosmology of interest. A workaround to this problem is to run a single time a relatively large number of simulations to sample the parameter space of interest and use efficient high-dimensional interpolation techniques to approximate the output of the simulations (in this case the matter power spectrum) in a dense region of the parameter space. This is for example what was done for the Euclid Emulator 2 (EE2) \cite{Euclid:2020rfv}, a state-of-the-art cosmological emulator for the matter power spectrum, which was claimed to be percent level accurate in a wide region of the cosmological parameter space. The EE2 relies on polynomial chaos expansion \cite{PCE} to approximate the output of its \textit{training} simulations. To simplify the interpolation problem, instead of using the full matter power spectrum as target function, the EE2 uses the non-linear boost factor of the power spectrum, i.e.\,the ratio of the non-linear power spectrum computed from simulations with the linear power spectrum computed from the Boltzmann solver CLASS~\cite{Diego_Blas_2011}. Therefore predicting the power spectrum with EE2 requires the evaluation of the CLASS code, which, while much faster than cosmological simulations, still requires about a second to run. In this work we are more interested in having a faster evaluation time than a high accuracy, so we use the power spectrum computed with EE2 in combination with CLASS as target to create a symbolic approximation which can be computed in microseconds, while providing $\sim 10 \%$ accurate power spectra. 

Our approach consists in spanning a subset of the EE2 parameter space in the variables $h$, $k$, $z$, while fixing the other parameters to the Planck 2018 fiducial cosmology~\cite{Planck:2018vyg}, as summarized in table~\ref{tab:param_space}.
\begin{table}[]
    \centering
    \begin{tabular}{lc}
        \toprule
        $h$  & $[0.61, 0.73]$ \\
        $k \, [\hompc]$ & $[0.001, 9.4]$\\
        $z$ & $[0, 1]$\\
        \midrule
        $\Omega_{\rm m}$  & $0.3111$ \\
        $\Omega_{\rm b}$  & $0.04897$ \\
        $m_{\nu}\, [{\rm eV}]$  & $0.06\,$ \\
        $n_{\rm s}$  & $0.9667$ \\
        $A_{\rm s}$  & $2.215\times 10^{-9}$ \\
        \bottomrule		
    \end{tabular}
    \caption{Ranges and fiducial values of the cosmological parameters used to create the training set for the approximate matter power spectrum expression~\eqref{eq:Power_Spectrum}.}
    \label{tab:param_space}
\end{table}%
We use a latin hypercube sampling (LHS) technique~\cite{LatinHypercubeSampling} to uniformly sample the ($h, k, z$) parameter space with 300 points and apply the symbolic regression algorithm \texttt{pyoperon}~\cite{10.1145/3377929.3398099} to obtain candidate symbolic expressions of various complexities, following an approach similar to the one outlined in~\cite{Bartlett:2024jes}. We select the model of least complexity that achieves better than $10\%$ accuracy\footnote{Unlike the common practice in literature of quoting the root mean squared error (RMSE) for the accuracy, we use here the maximum error on the training set, which we find being $\sim 3$ times larger than the RMSE.} on an independent test set of 1000 points, also constructed with LHS.

\bibliographystyle{JHEP}
\bibliography{references}

\end{document}

%% file: macros.tex
\def \be {\begin{equation}}
\def \ee {\end{equation}}
\def \dd {\mathrm{d}} 
\def \t {\tilde}

\def \l {\left}
\def \r {\right}

\def \bs {\boldsymbol}

\newcommand{\e}[1]{_{\rm #1}}

\newcommand{\beq}{\begin{equation}}
\newcommand{\eeq}{\end{equation}}
\newcommand{\bea}{\begin{eqnarray}}
\newcommand{\eea}{\end{eqnarray}}


\def\dd{\mathrm{d}}

\renewcommand\S{\mathcal{S}}

\newcommand\ees{\end{eqnarray}}
\newcommand\bees{\begin{eqnarray}}

\newcommand{\hompc}{\,h\,{\rm Mpc}^{-1}}